# Artificially creating emergent interfacial antiferromagnetism and its manipulation in a magnetic van-der-Waals heterostructure

*Xiangqi Wang*[1†], *Cong Wang*[2,3†], *Yupeng Wang*[4†], *Chunhui Ye*[4], *Azizur Rahman*[4], *Min Zhang*[5], *Suhan Son*[6,7,8¶], *Jun Tan*[1★], *Zengming Zhang*[4★], *Wei Ji*[2,3★], *Je-Geun Park*[6,7,8], *and Kai-Xuan Zhang*[6,7,8†★]

[1]Jihua Laboratory Testing Center, Jihua Laboratory, Foshan 528000, China

[2]Beijing Key Laboratory of Optoelectronic Functional Materials & Micro-Nano Devices, School of Physics, Renmin University of China, Beijing 100872, China

[3]Key Laboratory of Quantum State Construction and Manipulation (Ministry of Education), Renmin University of China, Beijing 100872, China

[4]Deep Space Exploration Laboratory, The Centre for Physical Experiments and CAS Key Laboratory of Strongly Coupled Quantum Matter Physics, Department of Physics, School of Physical Sciences, University of Science and Technology of China, Hefei 230026, China

[5]Institutes of Physical Science and Information Technology, Anhui University, Hefei 230601, China

[6]Department of Physics and Astronomy, Seoul National University, Seoul 08826, South Korea

[7]Center for Quantum Materials, Department of Physics and Astronomy, Seoul National University,

Seoul 08826, South Korea

[8]Institute of Applied Physics, Seoul National University, Seoul 08826, South Korea

[¶]Present address: Department of Physics, University of Michigan, Ann Arbor, MI 48109, USA

[†] These authors contributed equally to this work.

[★] Corresponding authors: Jun Tan (email: tanjun@jihualab.ac.cn), Zengming Zhang (email: zzm@ustc.edu.cn), Wei Ji (email: wji@ruc.edu.cn) and Kai-Xuan Zhang (email: kxzhang.research@gmail.com).

**ABSTRACT:** Van der Waals (vdW) magnets, with their two-dimensional (2D) atomic structures, provide a unique platform for exploring magnetism at the nanoscale. Although there have been numerous reports on their diverse quantum properties, the emergent interfacial magnetism—artificially created at the interface between two layered magnets—remains largely unexplored. This work presents observations of such emergent interfacial magnetism at the ferromagnet/antiferromagnet interface in a vdW heterostructure. We report the discovery of an intermediate Hall resistance plateau in the anomalous Hall loop, indicative of emergent interfacial antiferromagnetism fostered by the heterointerface. This plateau can be stabilized and further manipulated under varying pressures but collapses under high pressures over 10 GPa. Our theoretical calculations reveal that charge transfer at the interface is pivotal in establishing the interlayer antiferromagnetic spin-exchange interaction. This work illuminates the previously unexplored emergent interfacial magnetism at a vdW interface comprised of a ferromagnetic metal and an antiferromagnetic insulator, and highlights its gradual evolution under increasing pressure. These findings enrich the portfolio of emergent interfacial magnetism and support further investigations on vdW magnetic interfaces and the development of next-generation spintronic devices.

**KEYWORDS:** Artificially emergent interfacial antiferromagnetism; van-der-Waals heterostructure; ferromagnet/antiferromagnet heterostructure; intermediate Hall resistance plateau; advanced high-pressure nanotransport technique; interface charge transfer and magnetism.

In recent years, magnetic van-der-Waals (vdW) materials [1-6] have emerged as a focal point of intense research, captivating the global scientific community. These layered magnetic materials provide a unique opportunity to investigate and realize a range of fundamental quantum phenomena in unprecedentedly true two-dimensional (2D) systems [2,4,7-9], and introduce intriguing concepts in spintronics such as the gigantic spin filter effect [10-12] and intrinsic spin-orbit torque [13-15], significantly advancing magnetic device applications [10-24]. Besides, vdW magnets naturally facilitate the assembly of multifunctional heterostructures, leveraging their vdW nature and unique magnetic features. Despite the numerous studies on vdW magnetic heterostructures, the creation of emergent interfacial magnetism through proximity effects remains elusive. Furthermore, the manipulation of such emergent interfacial magnetism via external stimuli such as high pressure is even less explored and represents a frontier in this field.

$Fe_3GeTe_2$ (FGT) [25,26] represents the first studied vdW ferromagnetic metal of perpendicular magnetization with rich intriguing properties [13-15,27-33]. Extensive research efforts have focused on tuning the magnetic properties of FGT through diverse methods [26,34-40], but the creation of emergent magnetism by interfacing FGT with another vdW magnet is missing. To explore the possibility of emergent interfacial magnetism, we selected the antiferromagnet $MnPS_3$ (MPS) to form a heterostructure with FGT (FGT/MPS) for two primary reasons: First, MPS is an insulating antiferromagnet [41], which prevents the influence of free charge transport or magnetic stray fields, thereby isolating the interfacial effects. Second, as a representative Heisenberg model magnet[1], MPS exhibits relatively isotropic exchange interactions along different directions, which can accommodate any potentially required exchange interactions at the heterointerface. With the FGT/MPS heterostructure, we employed advanced high-pressure nanotransport techniques to manipulate and detect the emergent interfacial magnetism. Distinct from the high-pressure studies on bulk vdW magnets, the high-pressure nanotransport of nanoscale thin-layers and atomic interface counterparts has been rarely reported, which is both conceptually and technically challenging.

Except for the above possible emergent interfacial magnetism, another significant contribution of our current work lies in elucidating the effect of interfacial charge transfer. In principle, interfacial charge transfer exists in a heterointerface generally and universally. However, the recognition and understanding of its modulation effect on the interfacial proximal interlayer magnetic coupling is rather lacking. Our experiments, combined with theory, aim to demonstrate this modulation effect through the ferromagnet/antiferromagnet FGT/MPS heterostructure. Most importantly, this interfacial charge transfer and its modulation on interlayer magnetic coupling shall be a general principle applicable to all heterointerfaces beyond our specific investigated case. This fundamental mechanism, though critical, has been largely overlooked, particularly in the field of 2D vdW magnets. It opens up exciting possibilities for manipulating magnetism and developing corresponding quantum devices based on low-dimensional heterointerfaces.

In this work, we explore the interfacial effects within a vdW ferromagnet/antiferromagnet FGT/MPS heterostructure. We observe the development of an intermediate Hall resistance plateau during ferromagnetic switching induced by an external magnetic field in the heterostructure, which contrasts sharply with the abrupt switching observed in bare FGT nanoflakes. This plateau indicates the formation of an emergent antiferromagnetic phase near the heterostructure interface, referred to as homointerlayer antiferromagnetism induced by the heterointerface. Furthermore, we find that the antiferromagnetic plateau region expands under high pressure, with the critical onset magnetic field required to disrupt the antiferromagnetic plateau increasing correspondingly. This stabilization of the emergent interfacial antiferromagnetic phase under pressure is likely due to the enhanced interfacial interactions, as the high pressure brings the FGT and MPS layers closer together. However, when the pressure exceeds 10 GPa, FGT undergoes a transition from a 2D to a 3D-like structure. Such a 2D-to-3D crossover can suppress the interfacial interactions between the MPS and the 3D-like FGT compared to the 2D FGT, resulting in a return to the single ferromagnetic switching behavior characteristic of the anomalous Hall effect without an intermediate plateau. Our theoretical calculations highlight the critical role of charge transfer at the interface in generating interfacial antiferromagnetic exchange interactions. These findings reveal the surprising emergence of interfacial antiferromagnetism and its comprehensive modulation under high pressure, facilitated by our newly developed advanced high-pressure nanotransport technique. Our work not only elucidates the magnetic proximity effect and interfacial magnetism but also advances potential device applications in spintronics by utilizing the rapidly growing field of magnetic vdW materials.

**RESULTS AND DISCUSSIONS**

**Effect of pressure on a bare FGT nanoflake.** As shown in Fig. 1a, we developed a high-pressure nanotransport technique to measure the electrical transport properties of small-sized nanoflake devices under high pressure. Figure 1b displays a typical optical image of the FGT nanoflake device, with a thickness of approximately 25.4 nm, confirmed by atomic force microscopy. Figure S1-2 summarizes the transverse Hall resistance $R_{xy}$ as a function of the applied out-of-plane magnetic field under various pressures. To understand the magnetic evolution under pressure, we extracted key magnetic parameters, namely, the remnant Hall resistance $R_{xy}^r$ and coercivity $H_c$ at each temperature and pressure. As depicted in Fig. 1c, $R_{xy}^r$ decreases with increasing pressure until it vanishes at a fixed temperature (upper panel). Similarly, $R_{xy}^r$ decreases with increasing temperature at a fixed pressure (lower panel), which is consistent with the typical temperature dependence of magnetization in ferromagnets.

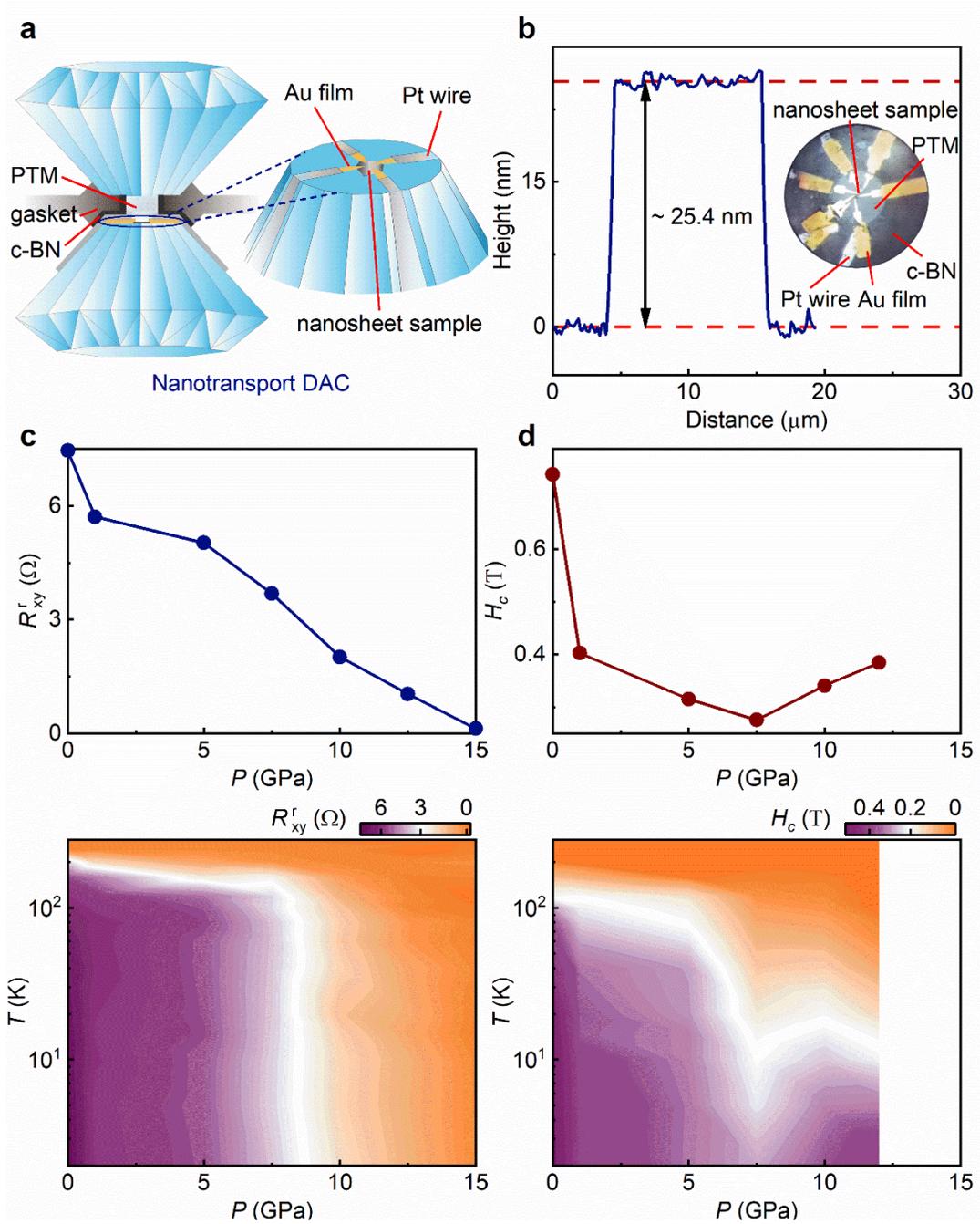

**Fig. 1. High-pressure modulation of nanotransport properties on a bare FGT nanoflake**

**device.** (a) 3D illustration of high-pressure nanotransport measurements. (b) Optical image of the bare FGT nanoflake device. The thickness was confirmed to be about 25.4 nm by the atomic force microscopy. (c) Pressure-dependent remnant Hall resistance $R_{xy}^r$ at 2 K. $R_{xy}^r$ shows a gradually decreasing trend with increasing pressure, which is consistent with that of bulk FGT under pressure in our previous work[36]. The lower panel represents the temperature-pressure mapping of $R_{xy}^r$. (d) Pressure-dependent coercivity $H_c$ at 2 K. $H_c$ decreases below 10 GPa but increases above 10 GPa as the pressure increases. The lower panel represents the temperature–pressure mapping of $H_c$.

Figure 1d illustrates the pressure and temperature dependence of coercivity $H_c$: both increasing pressure and temperature gradually suppress $H_c$. However, at extremely high pressures above 10 GPa, $H_c$ starts to increase with pressure, indicating underlying changes, such as a 2D to 3D crossover under high pressure. There are at least two pieces of key evidence to support such a 2D-to-3D crossover of FGT: First, previous Raman studies have shown a dynamical instability of the lattice [42] around 10 GPa, which can lead to structural transitions or a 2D to 3D-like crossover in structural, electronic, or magnetic properties frequently observed in highly-pressurized vdW systems [43-46]. Second, our X-ray diffraction measurements of FGT reveal that the unit cell volume decreases with increasing pressure but shows a slower decreasing trend above 10 GPa with a noticeable kink (Fig. S3).

The high-pressure nanotransport experiment on the FGT nanoflake provides more detailed information beyond high-pressure measurements on bulk FGT, such as the evolution of $H_c$, since the coercivity of bulk FGT is negligible. Nonetheless, the general pressure-induced magnetic evolution of bare FGT nanoflakes is similar to that of bulk FGT reported in our previous study [36], validating our newly developed high-pressure nanotransport technique for nanodevices.

**Emergent interfacial magnetism in the FGT/MPS heterostructure.** After examining the properties of bare FGT nanoflakes under high pressure, we now focus on the magnetic proximity effect in the FGT/MPS heterostructure. Figure 2a presents a schematic diagram of the crystal structure of the FGT/MPS heterostructure, and Fig. 2b provides the thickness information of a real device, with FGT at 7.7 nm and MPS at 8.9 nm. As MPS was intentionally selected due to its insulating properties and Heisenberg antiferromagnetism, the $R_{xy}-H$ curve of the heterostructure at zero pressure closely resembles that of pristine bare FGT nanoflakes at all temperatures (Fig. 2c), primarily reflecting the transport behavior of the magnetic FGT. However, in stark contrast to the direct and smooth switching of bare FGT, the magnetic switching of the FGT/MPS heterostructure exhibits an intermediate Hall resistance plateau, as indicated by the dashed black arrow.

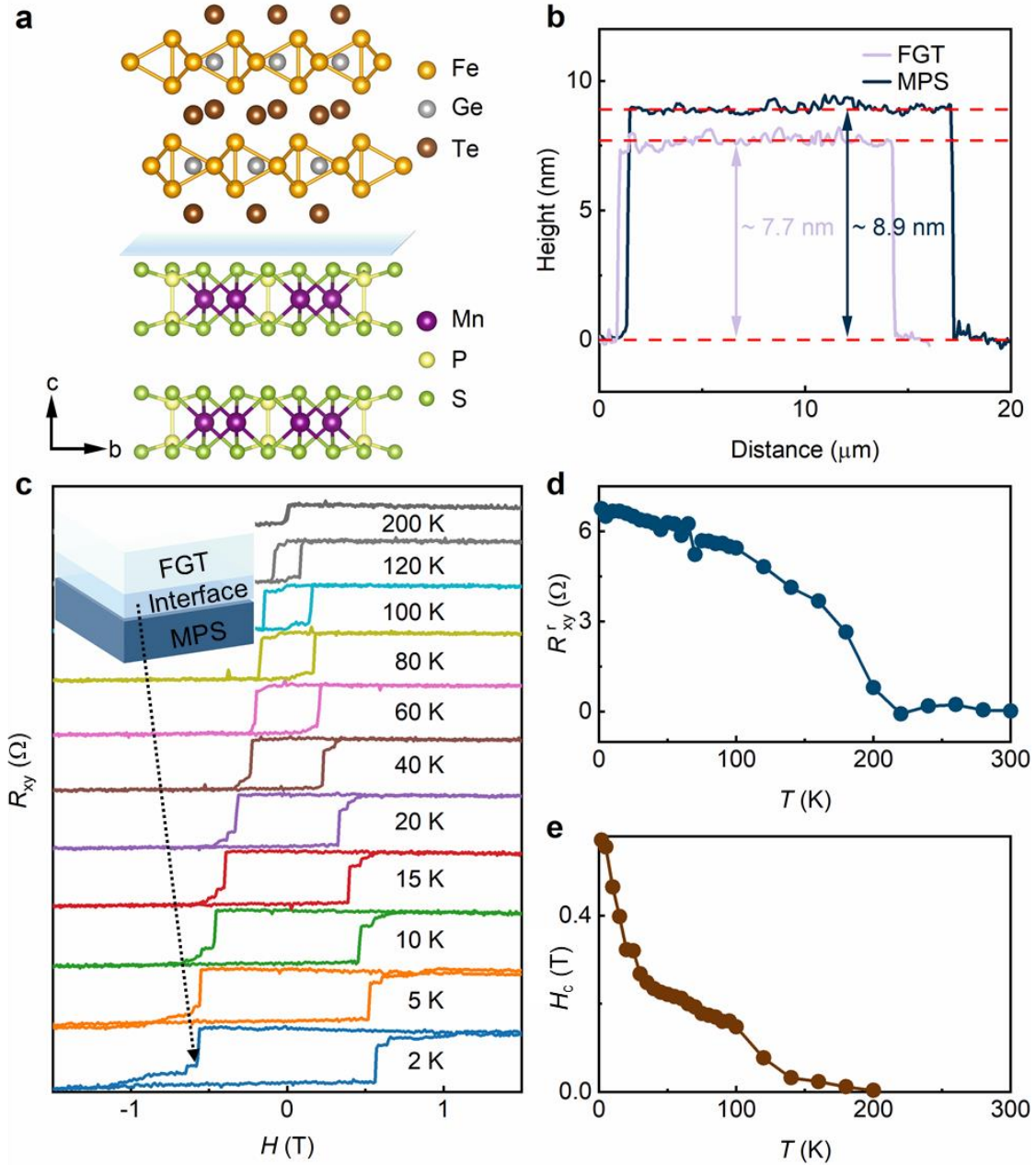

**Fig. 2. Emergent interfacial magnetism in the FGT/MPS heterostructure.** (a) Illustration of the FGT/MPS heterostructure. (b) Thickness information of a typical real nanodevice with an FGT (7.7 nm)/MPS (8.9 nm) heterostructure. (c) $R_{xy}$-$H$ curves under zero pressure at various temperatures while sweeping the out-of-plane magnetic field $H$. The dashed black arrow highlights the intermediate Hall resistance plateau, which indicates emergent interfacial magnetism in the device. Such intermediate Hall resistance appears both in the positive and negative magnetic field

direction symmetrically. The transport device's contribution is described by the three-layer toy model in the inset: the pristine FGT part and the interfacial magnetism part. (d-e) $R_{xy}^r$ and $H_c$ as a function of temperature $T$. They gradually decrease with increasing temperature, featuring a normal evolution of a typical ferromagnet FGT, except for the intermediate Hall plateau.

This emergent Hall plateau is a unique feature of the FGT/MPS hybrid system and is most prominent at low temperatures between 2 and 20 K, where both FGT and MPS maintain their ferromagnetic and antiferromagnetic ground states, respectively. This observation suggests the presence of distinct characteristics in the heterostructure, which can be attributed to the interfacial pinning effect, intrinsically originated from an emergent antiferromagnetism by the vdW heterointerface. This phenomenon is demonstrated in our case of a ferromagnetic-metal/antiferromagnetic-insulator interface. To illustrate this, we included a schematic inset showing three parts: FGT, the interface, and the insulating MPS, to describe the transport contributions from both the FGT and the interfacial region within the entire device. Apart from this difference, all other features, such as the temperature-dependent variations in $R_{xy}^r$ and $H_c$, are similar to the typical ferromagnetic behavior of pristine FGT (Fig. 2d,e).

**Pressure modulation of emergent interfacial magnetism.** To better understand the emergent interfacial magnetism, we performed high-pressure nanotransport measurements on the FGT/MPS heterostructure. Figure 3a presents the $R_{xy}$-$H$ curves of the FGT/MPS heterostructure at 2 K under various pressures. The dashed black lines highlight $H_c$ and $H_I$, representing the coercivity of the FGT component and the critical onset magnetic field required to disrupt the interfacial magnetism, respectively. As shown in Fig. 3b,c, $R_{xy}^r$ decreases with increasing pressure, and $H_c$ decreases below 10 GPa but increases above 10 GPa, which is consistent with the behavior observed in bare FGT nanoflakes (Fig. 1) and bulk FGT in our previous study [36], reflecting the intrinsic properties of pristine FGT. In sharp contrast, $H_I$ and $H_I$-$H_c$ increase with increasing pressure (Fig. 3d), which cannot be attributed to the FGT component but rather to the emergent interfacial magnetism. This finding indicates that the interfacial magnetism is further stabilized by increasing pressure, which contrasts with the pressure-weakened ferromagnetism of FGT and suggests the antiferromagnetic nature of the interfacial magnetism. Applying higher pressure brings MPS into greater proximity to FGT, thereby enhancing the interfacial interaction. This increased interfacial interaction aligns with the pressure-stabilized intermediate Hall plateau, further corroborating that the plateau originates from emergent interfacial antiferromagnetism within the heterostructure.

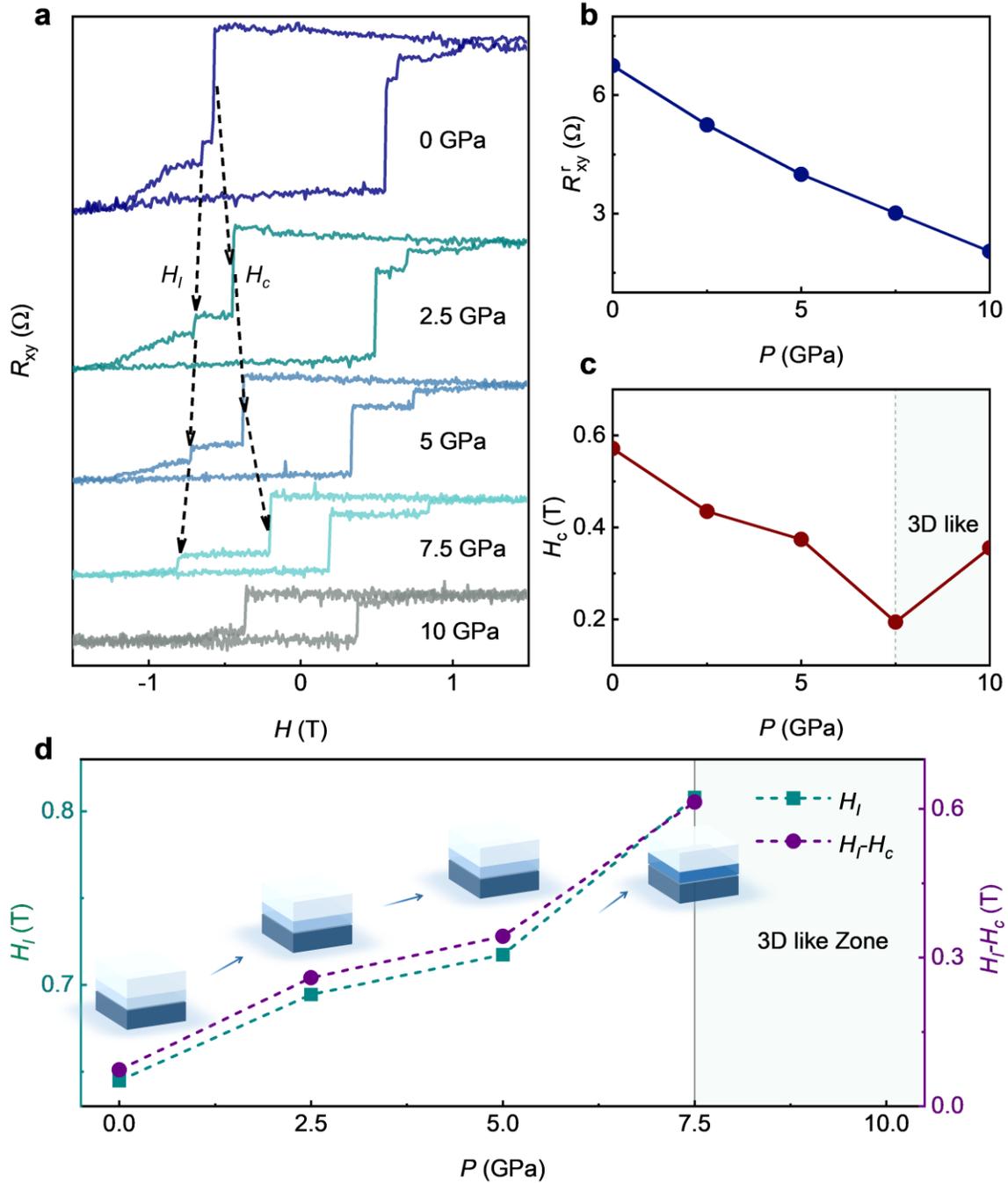

**Fig. 3. High-pressure tuning of the emergent interfacial magnetism.** (a) $R_{xy}$-$H$ curves at 2 K under different pressures, e.g., 0, 2.5, 5, 7.5, and 10 GPa. The dashed black arrows highlight the pressure-dependent evolution of $H_I$ and $H_c$, which represents the onset field for destruction of the interfacial magnetism and the coercivity of the FGT part. (b-c) $R_{xy}^r$ and $H_c$ of the pristine FGT part

as a function of pressure in (a). They show a normal reduction trend, similar to that of the bare FGT nanoflakes in Fig. 1 and the bulk FGT in our previous work [36]. (d) $H_I$ and $H_I$-$H_c$ as a function of applied pressure. Both $H_I$ and $H_I$-$H_c$ increase with increasing pressure, indicating enhanced stabilization of the emergent interfacial magnetism under high pressure. At the final extreme pressure, the intermediate plateau disappears, and the magnetic switching behaves uniformly and directly at 10 GPa. The magnetic evolution is also illustrated correspondingly with the inset schematic. The interfacial magnetism is represented by the light blue box, which becomes stronger as the pressure increases.

It is noteworthy that MPS well preserves its insulating and antiferromagnetic characteristics below 10 GPa [47-49]. However, at approximately 10 GPa, FGT undergoes a 2D to 3D-like crossover, which significantly suppresses the interfacial interaction between MPS and 3D-FGT. Consequently, the magnetic switching becomes uniform, without interfacial contributions (Fig. 3, a and d). On the basis of these analyses, we constructed a schematic of the magnetism evolution in the heterostructure under pressure, as shown in the inset of Fig. 3d. As the pressure increases, the ferromagnetism of the FGT component (indicated in white) gradually weakens, whereas the emergent interfacial antiferromagnetism at the FGT/MPS interface (indicated in light blue) increasingly stabilizes. At extreme pressure, the interfacial effect vanishes due to the 2D to 3D-like crossover of FGT, resulting in uniform and direct magnetic switching.

**Temperature and pressure mapping of the interfacial magnetism.** To provide a comprehensive overview of the evolution of interfacial magnetism, we performed high-pressure measurements of $R_{xy}$-$H$ curves across various temperatures. Figure 4a-e displays the temperature–magnetic field maps of the normalized Hall resistance $R_{xy}^n$ at 0, 2.5, 5, 7.5, and 10 GPa, respectively. Here, $R_{xy}^n$ is defined as $R_{xy}/\max(R_{xy})$ at different magnetic fields $H$. In these mappings, $H_c$ and $H_I$ are highlighted and increase as the system temperature decreases. However, $H_I$ increases much more rapidly than $H_c$ does with decreasing temperature, particularly at higher pressures. Notably, $H_c$ shows a suppressed rate of increase with decreasing temperature as the pressure increases, whereas $H_I$ evolves in the opposite direction. This results in a larger discrepancy between $H_I$ and $H_c$, i.e., $H_I$-$H_c$ increases under higher pressures. This pressure-dependent behavior is similar to the results observed at the lowest temperature of 2 K in Fig. 3. At a high pressure of 10 GPa, the interfacial magnetism disappears, and the $H_c$ of the uniform magnetic switching starts to increase compared with that at 7.5 GPa, which is also consistent with the pressure-dependent evolution at 2 K shown in Fig. 3. More importantly, we plotted the temperature–pressure maps of $H_I$ and $H_I$-$H_c$ in Fig. 4f by extracting data from Fig. 4a-e. As shown, both $H_I$ and $H_I$-$H_c$ increase as the temperature decreases or the pressure increases. In other words, the interfacial magnetism can be modulated by both temperature and pressure and is most pronounced in the low-temperature–high-pressure regime, as indicated by the reddest region in the bottom-right corner of the map.

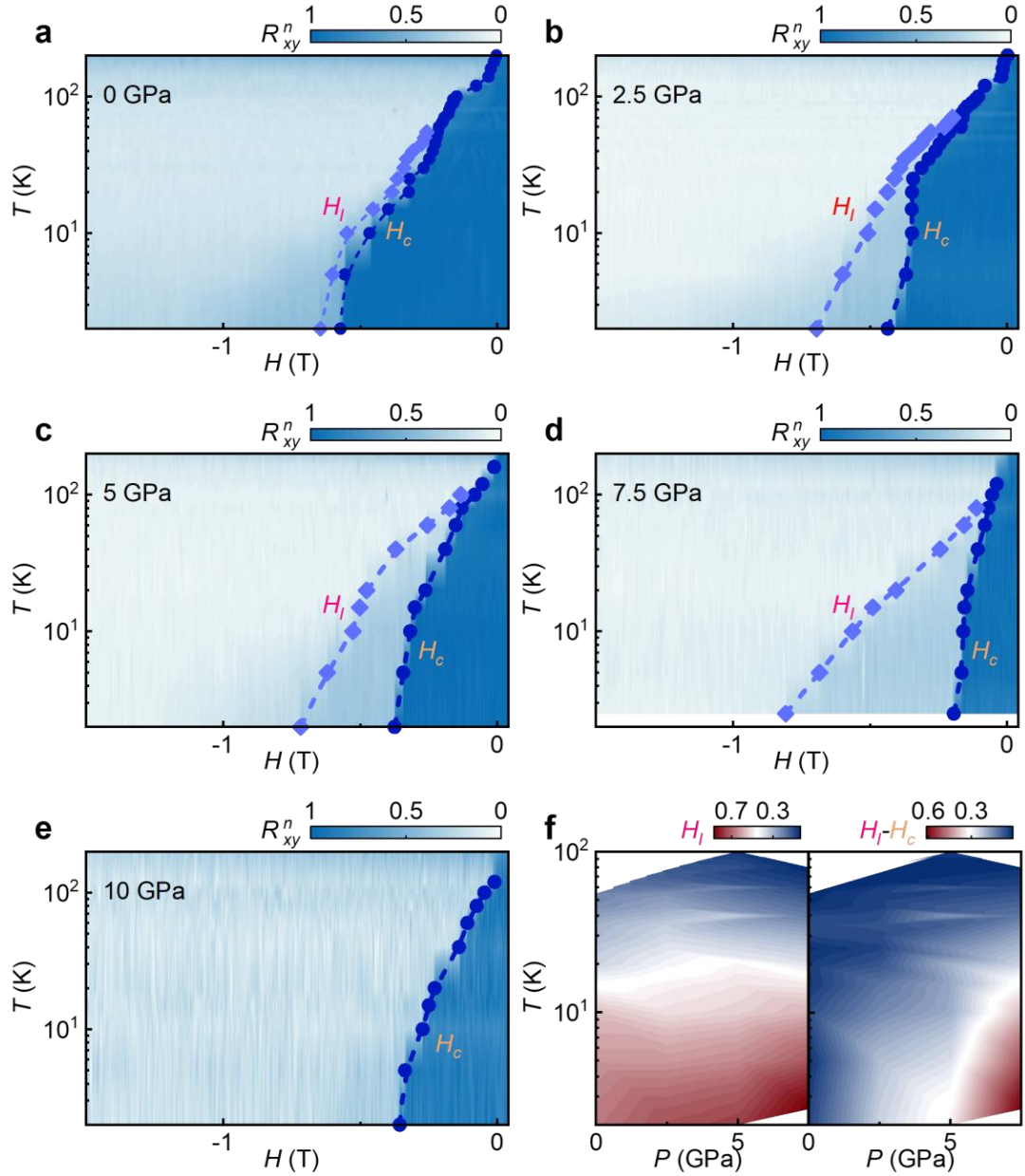

**Fig. 4. Temperature and pressure mapping of magnetism.** (a-e) Temperature–magnetic field mapping of $R_{xy}^n$ (= $R_{xy}/\max(R_{xy})$) for individual pressures, e.g., 0 (a), 2.5 (b), 5 (c), 7.5 (d), and 10 (e) GPa, respectively. $H_I$ and $H_c$ are highlighted correspondingly. (f) Temperature–pressure mapping of $H_I$ and $H_I$-$H_c$, derived from (a-e). Note that both $H_I$ and $H_I$-$H_c$ increase at lower temperatures and higher pressures, as does the interfacial magnetism, which is mostly strengthened at the right-bottom corner with the reddest color.

**Theoretical calculations.** Density functional theory (DFT) calculations were performed to elucidate the substrate-modulated magnetism in few-layer FGT. A $3\times3\sqrt{3}$ 3L-FGT/$2\times2\sqrt{3}$ BL-MPS supercell (Fig. S7) was constructed to model the experimentally obtained FGT/MPS heterostructure, which corresponds to substrate-induced compressive in-plane strains of 2.1 % (1.7 %) along the *x*-axis (*y*-axis) and lowest total energy within a fairly large search space. Our calculations reveal that while the substrate-induced interfacial strain weakens the interlayer ferromagnetic coupling of FGT, it does not render the antiferromagnetic state as the ground state (Table S1). Figure 5a,b illustrates charge accumulation upon stacking FGT on MPS, indicating that charge transfer predominately occurs between these interfacial layers of the heterostructure. Given the valence band edge at -6.2 eV relative to the vacuum level for MPS and a work function of 5.0 eV for FGT, electrons transfer from FGT to MPS in the FGT/MPS heterostructure, resulting in local hole doping to the interfacial FGT layer and thereby modifying the interfacial magnetic couplings.

In the free-standing trilayer FGT, the most energetically stable configuration features both intra- and interlayer ferromagnetic orders, (Fig. 5e) compared to other magnetic configurations (Fig. 5f,g) by at least 0.39 meV/Fe (Table S1). The local hole doping induced by the MPS substrate weakens the ferromagnetic interlayer coupling between adjacent FGT layers [26]. Consequently, an intralayer ferromagnetic and interlayer ferrimagnetic order (FM-AAB, Fig. 5g) becomes energetically favorable over the pure ferromagnetic order (Table S1). The interfacial interlayer antiferromagnetic coupling causes the interfacial and bulk FGT layers responding differently under an external field, potentially leading to the Hall resistance plateau observed in our experiments (Fig. 2).

To validate this mechanism, we introduced local hole doping into the lowest Te layer of the freestanding trilayer FGT to model the interfacial charge transfer. Our results show that the FM-AAB configuration becomes the magnetic ground state when the hole doping concentration exceeds 0.05 $h$/Te (Fig. 5h). Moreover, as the doping concentration increases from 0.05 to 0.225 $h$/Te, the relative stability of the FM-AAB state is further enhanced. The reduction in interfacial distance between FGT and MPS under external pressure increases the hole doping concentration (Fig. 5c,d), likely contributing to the observed enhancement of the coercive field $H_I$ under moderate pressure ranging from 2.5 to 7.5 GPa in our experiments. At 10 GPa, FGT, most likely, undergoes a transition from a 2D to a 3D-like structure [42], leading to a significant suppression of interfacial AFM magnetic coupling and causing the interface to become ferromagnetic as the ground state (Table S1), consistent with experimentally observed disappearance of the Hall resistance plateau.

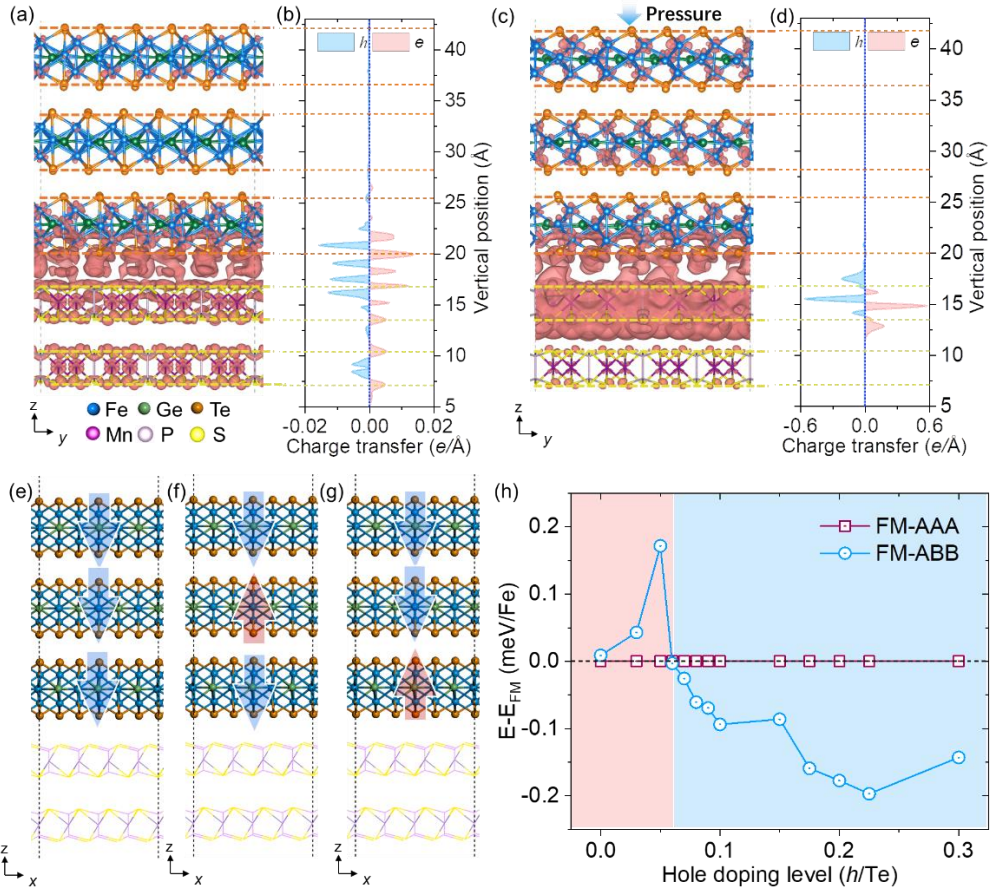

**Fig. 5. Substrate and pressure-modulated magnetism of few-layer FGT.** (a-d) Hetero-interlayer differential charge density (a, c) and corresponding line profiles along *z* (b, d) of the FGT trilayer stacked on the MPS bilayer without (a, b) and with (c, d) vertical strain. An isosurface value of 0.001 $e\,\mathrm{Bohr}^{-3}$ was used. The red isosurface contours represent charge accumulation after layer stacking. The blue, green, brown, purple, pink and yellow balls represent the Fe atoms, Ge atoms, Te atoms, Mn atoms, P atoms and S atoms, respectively. (e-g) Schematic representations of interlayer magnetic orders, including FM-AAA (e), FM-ABA (f) and FM-ABB (g) are depicted. The yellow and purple lines correspond to the MPS substrate for clarity. The blue and red arrows represent two antiparallel orientations of magnetic moments on Fe atoms. h) Relative total energies of the FM-AAA (red rectangles) and FM-AAB (blue circles) orders of FGT trilayer as a function of local hole doping on bottom Te atoms. The FM-AAA order was chosen as the reference zero.

# CONCLUSIONS

In summary, we have artificially induced emergent interfacial magnetism at the vdW magnetic interface in the FGT/MPS heterostructure. This emergent interfacial magnetism manifests as an intermediate Hall resistance plateau within the anomalous Hall effect and exhibits a markedly different evolutionary trend compared to ferromagnetic FGT under high pressures. Our theoretical analysis suggests that interfacial charge transfer may be responsible for the emergence and reinforcement of interfacial antiferromagnetism. Our work provides a valuable example of emergent interfacial magnetism and its manipulation, offering insights into interface magnetism engineering, 2D magnetic materials, and the development of advanced spintronic devices.

## METHODS

**Preparation of FGT, MPS single crystals, nanoflake devices, and FGT/MPS heterostructures**

High-quality FGT single crystals were grown *via* the chemical vapor transport technique [35,39]. A stoichiometric amount of Fe, Ge, and Te powder (Fe purity, 99.9%; Ge purity, 99.999%; Te purity, 99.999%) with a molar ratio of 3:1:2 was mixed and sealed in an evacuated quartz tube. Iodine was employed as a transport agent. The quartz tube was heated with a temperature gradient of 1023 K/973 K for several days. Thin flake crystals of millimeter size with metallic luster were obtained after natural cooling to room temperature. The crystals were kept in vacuum-sealed tubes to prevent possible oxidation.

MPS single crystals were also prepared via the chemical vapor transport method. A stoichiometric mixture of high-purity Mn, P, and S powders with a molar ratio of 1:1:3 was mixed and sealed in an evacuated quartz tube with iodine as a transport agent. The quartz tube was also heated with a temperature gradient of 1023 K to 973 K for several days. Plate-like MPS single crystals were obtained after natural cooling to room temperature.

Few-layer FGT and MPS samples were obtained by exfoliating bulk samples onto PDMS substrates. The heterostructures for transport measurements were fabricated by sequentially transferring FGT and MPS onto the diamond surface with prepatterned electrodes.

**High-pressure nanotransport measurements**

Nonmagnetic beryllium copper (BeCu) diamond anvil cells (DACs) with 500 μm culet type-2 diamonds were used for high-pressure experiments. Platinum (Pt) electrodes were positioned on a tungsten gasket, which was coated with a mixture of cubic boron nitride (c-BN) and epoxy for insulation. Daphne 7373 was employed as the pressure-transmitting medium, and the pressure was calibrated using ruby fluorescence.

After preparing the Pt electrodes on the gasket, their relative positions with respect to the diamond anvil were carefully recorded. Titanium/gold (Ti/Au, 5 nm/100 nm) electrodes were then patterned on the diamond surface using photolithography and deposited at the corresponding locations. This ensured reliable electrical connections between the FGT/MPS sample, placed onto the diamond, and the Pt electrodes.

For transport measurements, the Physical Property Measurement System (PPMS) was utilized, employing standard low-frequency lock-in amplifier techniques with an excitation current of 100 nA. This setup allowed for precise electrical characterization of the sample under high-pressure conditions.

**Computational methods**

The calculations were performed via the generalized gradient approximation and projector augmented-wave method [50] as implemented in the Vienna ab initio calculation package [51]. Dispersion correction was considered in Grimme's semiempirical D3 scheme [52] in combination with the PBE functional (PBE-D3). This combination achieves an accuracy comparable to that of the optB86b-vdW functional for describing the geometric properties of layered materials at a lower computational cost [53]. On-site Coulomb interactions on Fe and Mn 3$d$ orbitals are characterized via U and J values, namely, $U = 3.6$ eV and $J = 0.2$ eV for $Fe^{3+}$, $U = 4.6$ eV and $J = 1.0$ eV for $Fe^{2+}$ and U = 4.0 eV and J = 0 eV for $Mn^{2+}$, as revealed via a linear response method [37,54,55] via the PlusU package. A uniform Monkhorst–Pack $k$-mesh of size 15×15×3 was adopted for integrating over the Brillouin zone of an FGT bulk unit cell. The shape and volume of the unit cell of pristine FGT were fully optimized, and all the atoms were allowed to relax until the residual force per atom was less than 0.01 eV/Å. A 3×3$\sqrt{3}$ trilayer FGT/2×2$\sqrt{3}$ bilayer MPS magnetic supercell was adopted for the calculations of structure optimization and magnetic ground state (Fig. 5) to include the substrate effect, in which an MP $k$-mesh of 3 × 1 ×1 was used. A kinetic energy cutoff of 400 eV for the plane-wave basis set was used for both structural relaxation and electronic structure calculations of the FGT on MPS heterostructure models. A sufficiently large vacuum layer over 20 Å along the out-of-plane direction was adopted to eliminate the interaction among layers. Charge doping on Te atoms was realized with the ionic potential method[56,57]. For hole doping, electrons were removed from the valence band by adding a negative potential into the 4d core level of Te. This method ensured that the doped charges were located around the Te atoms. It also retained the neutrality of the whole supercell without introducing background charge, which eliminated the effects of compensating charges.

## AUTHOR INFORMATION

**Corresponding Authors: Jun Tan (email: tanjun@jihualab.ac.cn), Zengming Zhang (email: zzm@ustc.edu.cn), Wei Ji (email: wji@ruc.edu.cn) and Kai-Xuan Zhang (email: kxzhang.research@gmail.com).**

## Supporting Information

**Discussions on the interfacial magnetism of the FGT/MPS heterostructure; Anomalous Hall measurement of bare FGT under 0 GPa; $R_{xy}$-$H$ curves of the FGT nanoflake device at different temperatures under various pressures; Lattice constants and unit cell volume of FGT as a function of pressure; $R_{xx}$-$T$ of FGT under diverse pressures; $R_{xy}$-$H$ curves of the FGT/MPS heterostructure at different temperatures under various pressures.; $T$-$P$ mapping**

of $H_c$ in FGT/MPS heterostructure; Top and side views of the heterostructure model; Certain exchange bias effect in the samples; Relative total energies of the free-standing and substrate-supported 3L FGT.

ACKNOWLEDGMENT

X. W., C. W., Y. W., and K.-X. Z. contributed equally to this work. The work at Jihua Lab was supported by the GuangDong Basic and Applied Basic Research Foundation, No. 2021A1515110929, and the continuous scale preparation of electroscope samples and development of an in-situ characterization platform, No. X200191TL200. The work at USTC was supported by the National Natural Science Foundation of China (No. 12074360) and the Frontier Scientific Research Program of Deep Space Exploration Laboratory under grant No. 2022-QYKYJH-HXYF-019. The work at CQM and SNU was supported by the Samsung Science & Technology Foundation (Grant No. SSTF-BA2101-05), and one of the authors, Je-Geun Park, was funded by the Leading Researcher Program of the National Research Foundation of Korea (Grant No. 2020R1A3B2079375 and RS-2020-NR049405). The work at RUC was supported by the financial support from the Ministry of Science and Technology (MOST) of China (Grant No. 2023YFA1406500), the National Natural Science Foundation of China (Grants No. 92477205, 52461160327 and 12104504), the Strategic Priority Research Program of Chinese Academy of Sciences (Grant No. XDB30000000), the Fundamental Research Funds for the Central Universities, and the Research Funds of Renmin University of China [Grants No. 22XNKJ30 (W.J.) and 24XNKJ17 (C.W.)]. All calculations for this study were performed at the Physics Lab of High-Performance Computing (PLHPC) and the Public Computing Cloud (PCC) of Renmin University of China.

ToC figure

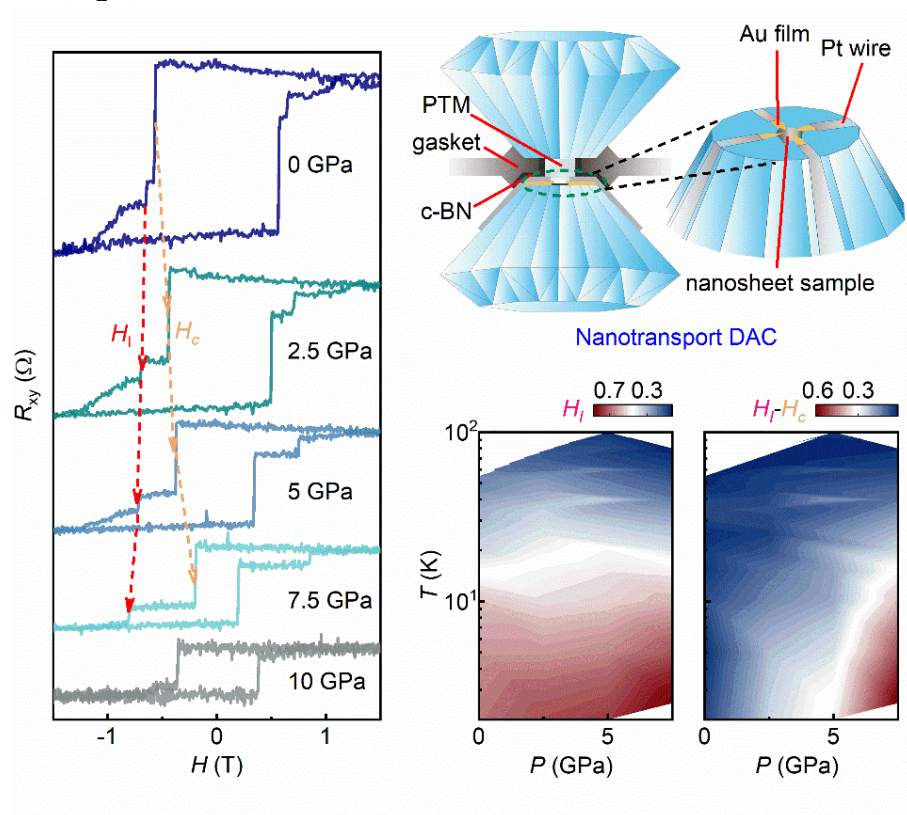

# Supporting Information for:

# Artificially creating emergent interfacial antiferromagnetism and its manipulation in a magnetic van-der-Waals heterostructure

*Xiangqi Wang*[1†], *Cong Wang*[2,3†], *Yupeng Wang*[4†], *Chunhui Ye*[4], *Azizur Rahman*[4], *Min Zhang*[5], *Suhan Son*[6,7,8¶], *Jun Tan*[1★], *Zengming Zhang*[4★], *Wei Ji*[2,3★], *Je-Geun Park*[6,7,8], and *Kai-Xuan Zhang*[6,7,8†★]

[1]Jihua Laboratory Testing Center, Jihua Laboratory, Foshan 528000, China

[2]Beijing Key Laboratory of Optoelectronic Functional Materials & Micro-Nano Devices, School of Physics, Renmin University of China, Beijing 100872, China

[3]Key Laboratory of Quantum State Construction and Manipulation (Ministry of Education), Renmin University of China, Beijing 100872, China

[4]Deep Space Exploration Laboratory, The Centre for Physical Experiments and CAS Key Laboratory of Strongly Coupled Quantum Matter Physics, Department of Physics, School of Physical Sciences, University of Science and Technology of China, Hefei 230026, China

[5]Institutes of Physical Science and Information Technology, Anhui University, Hefei 230601, China

[6]Department of Physics and Astronomy, Seoul National University, Seoul 08826, South

Korea

[7]Center for Quantum Materials, Department of Physics and Astronomy, Seoul National University, Seoul 08826, South Korea

[8]Institute of Applied Physics, Seoul National University, Seoul 08826, South Korea

[¶]Present address: Department of Physics, University of Michigan, Ann Arbor, MI 48109, USA

[†] These authors contributed equally to this work.

[★] Corresponding authors: Jun Tan (email: tanjun@jihualab.ac.cn), Zengming Zhang (email: zzm@ustc.edu.cn), Wei Ji (email: wji@ruc.edu.cn) and Kai-Xuan Zhang (email: kxzhang.research@gmail.com).

## Supporting Notes

**Note 1. Discussions on the interfacial magnetism of the FGT/MPS heterostructure**

We would like to discuss the interfacial magnetism and its pinning effect in the FGT/MPS heterostructure, i.e., the physical picture we proposed based on our experiment and theory:

(1) Our project was carefully designed to adopt the FGT/MPS heterostructure to investigate the vdW interface effect, as elaborated in the Introduction Section: First, MPS is an insulating antiferromagnet, which prevents the influence of free charge transport or magnetic stray fields, thereby isolating the interfacial effects. Second, as a representative Heisenberg model magnet, MPS exhibits relatively isotropic exchange interactions along different directions, which can accommodate any potentially required exchange interactions at the heterointerface. Meanwhile, FGT is a vdW ferromagnetic metal with typical out-of-plane Ising-type ferromagnetism with a significant anomalous Hall effect. Therefore, MPS/FGT is an ideal system to explore the unknown interface magnetic interaction of vdW magnet heterostructure, by using our newly developed advanced high-pressure nanotransport technique.

(2) Experimentally, we observe the intermediate Hall plateau in the AHE loop for the heterostructure, indicating that FGT's magnetization switching becomes harder than bare FGT. It is likely due to the general pinning effect of an interfacial magnetization.

(3) Moreover, applying an increasing pressure can continuously widen the intermediate Hall plateau and thus strengthen the interface magnetism, in sharp contrast to the weakened coercivity at increasing pressure for pristine FGT. It indicates that interface magnetism is distinct from the ferromagnetism of pristine FGT, showcasing instead an emergent interfacial antiferromagnetism.

(4) Furthermore, our theory reveals that the FGT layer nearest to the MPS is antiferromagnetically rather than ferromagnetically coupled with the farther FGT layers, and thus pins those farther layers to make them less switchable. Finally, an intermediate Hall plateau develops due to such interfacial magnetization pinning effects.

In summary, the anomaly in our FGT/MPS heterostructure comes from the interfacial magnetization pinning effect, which intrinsically originates from an emergent antiferromagnetism by the vdW heterointerface.

**Supporting Figures**

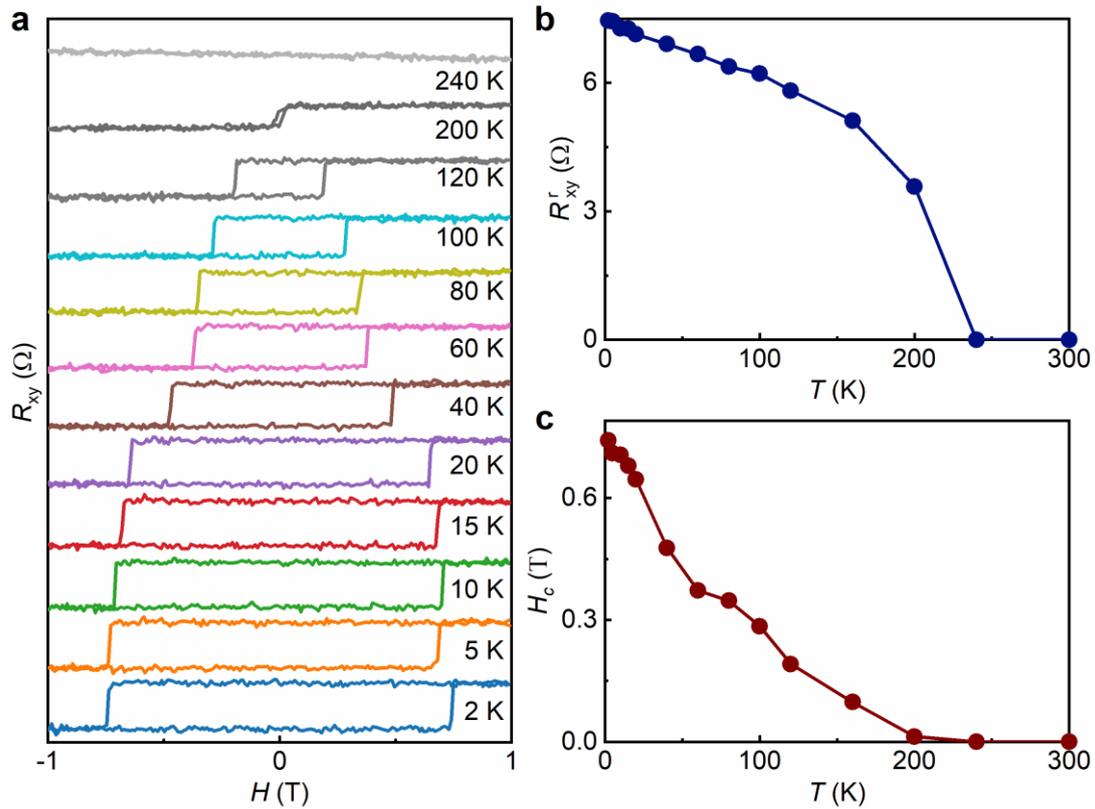

**Fig. S1. Anomalous Hall measurement of bare FGT under 0 GPa.** It shows the typical anomalous Hall effect with sharp rectangular hysteresis loops as reported in many previous investigations. This typical ferromagnetic hysteresis loop is in sharp contrast to the intermediate Hall plateau and its continuous pressure modulation of the heterostructure in our present work, highlighting the emergent magnetic behavior due to the vdW heterointerface.

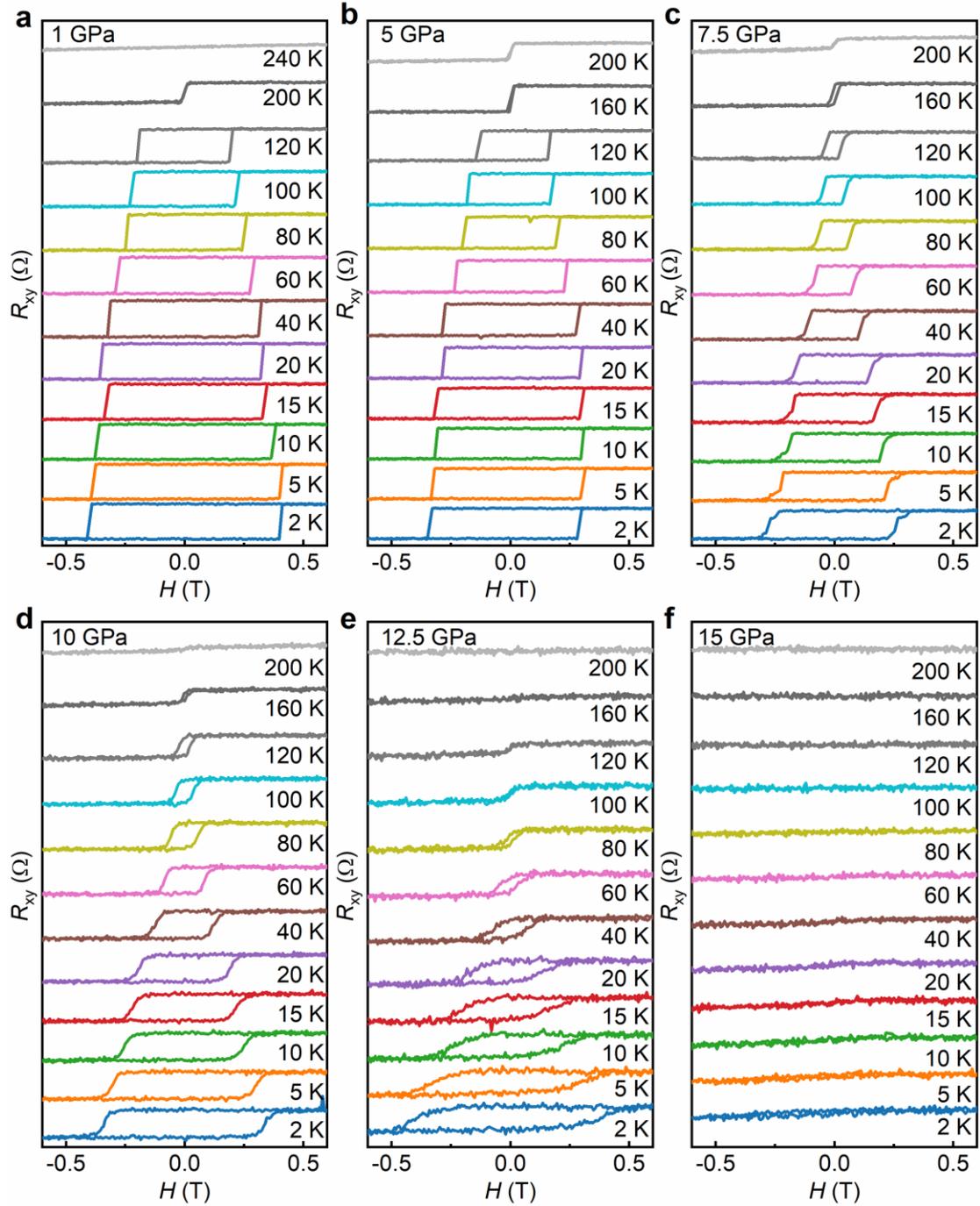

**Fig. S2. $R_{xy}$-H curves of the FGT nanoflake device at different temperatures under various pressures.** $R_{xy}$ represents the transverse Hall resistance and $H$ indicates the applied out-of-plane magnetic field. Each applied high pressure is 1 (a), 5 (b), 7.5 (c), 10 (d), 12.5 (e), and 15 (f) GPa, respectively.

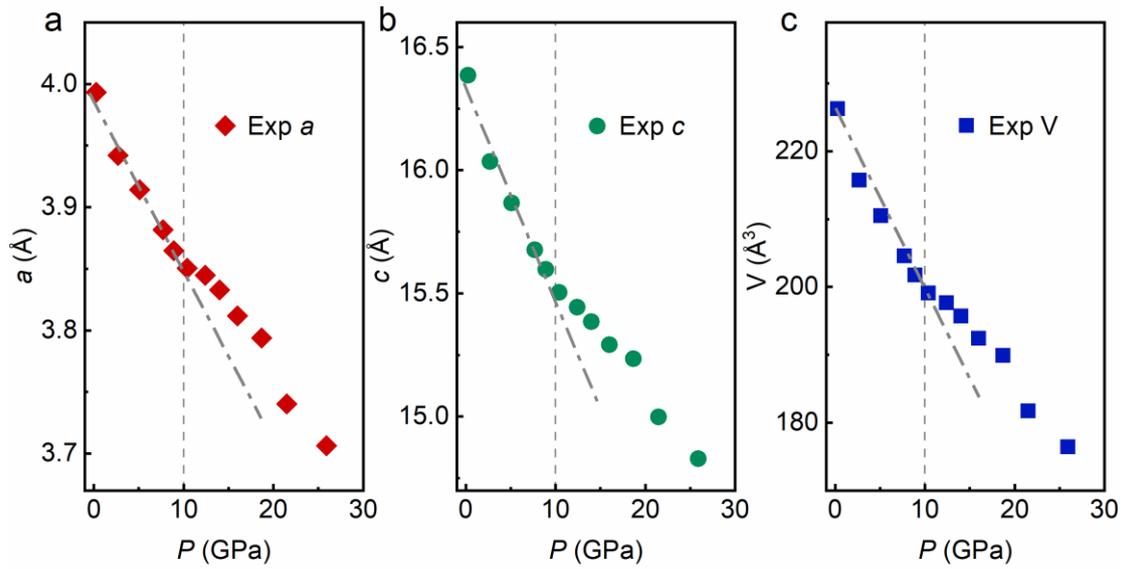

**Fig. S3. Lattice constants and unit cell volume as a function of pressure.** (a) Lattice constant *a*, (b) Lattice constant *c*, (c) unit cell volume V. The X-ray diffraction measurements of FGT reveal that the unit cell volume decreases with increasing pressure but shows a slower decreasing trend above 10 GPa with a noticeable kink. It indicates the possible 2D-to-3D-like crossover of FGT at around 10 GPa, consistent with previous Raman studies.

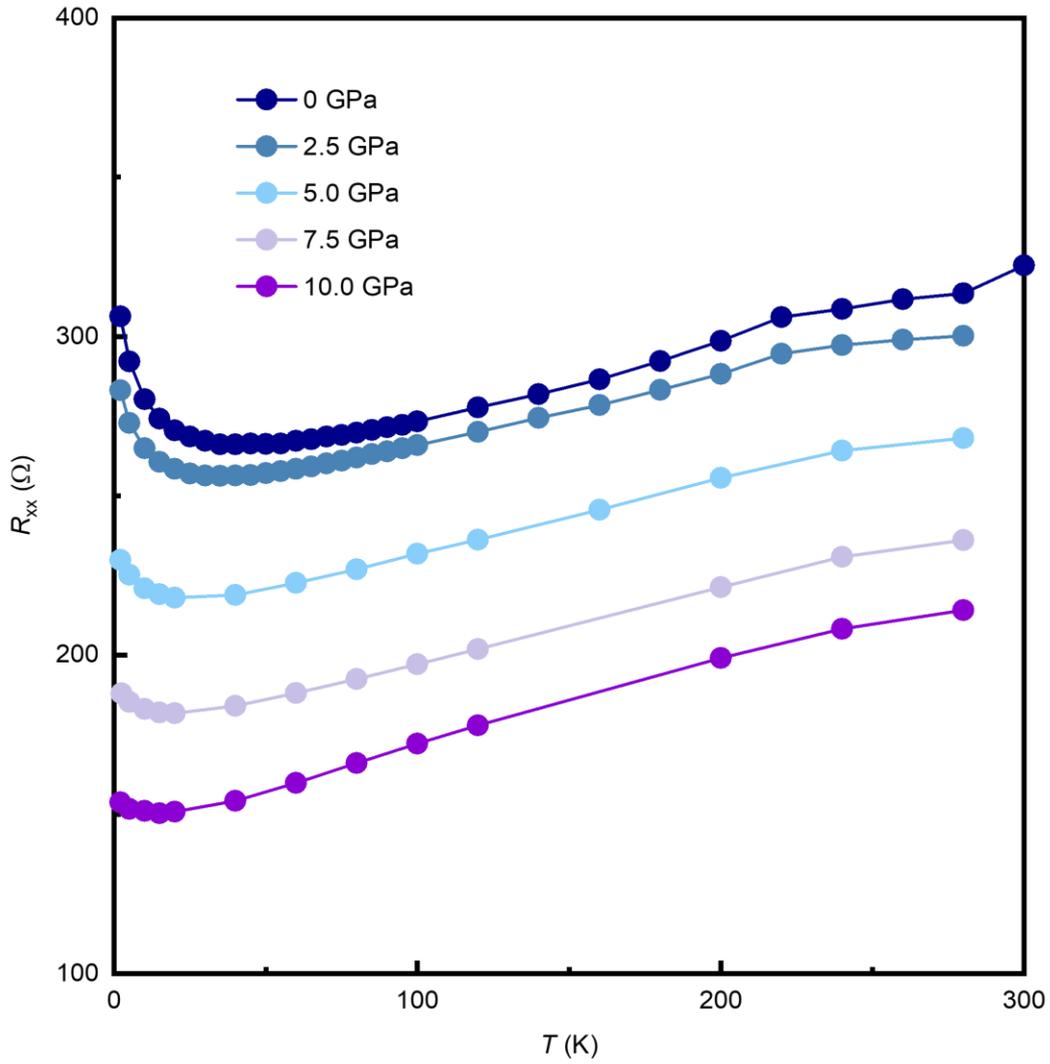

**Fig. S4. The temperature-dependent longitudinal resistance $R_{xx}$ of FGT under diverse pressures.** $R_{xx}$ shows a normal metallic behavior of decreasing values as temperature reduces and the kondo-like upturn in FGT. Note that the resistance $R_{xx}$ values decrease as pressure increases, which is natural metallization by high pressure in many materials, and thus validates the operation of applying high pressure.

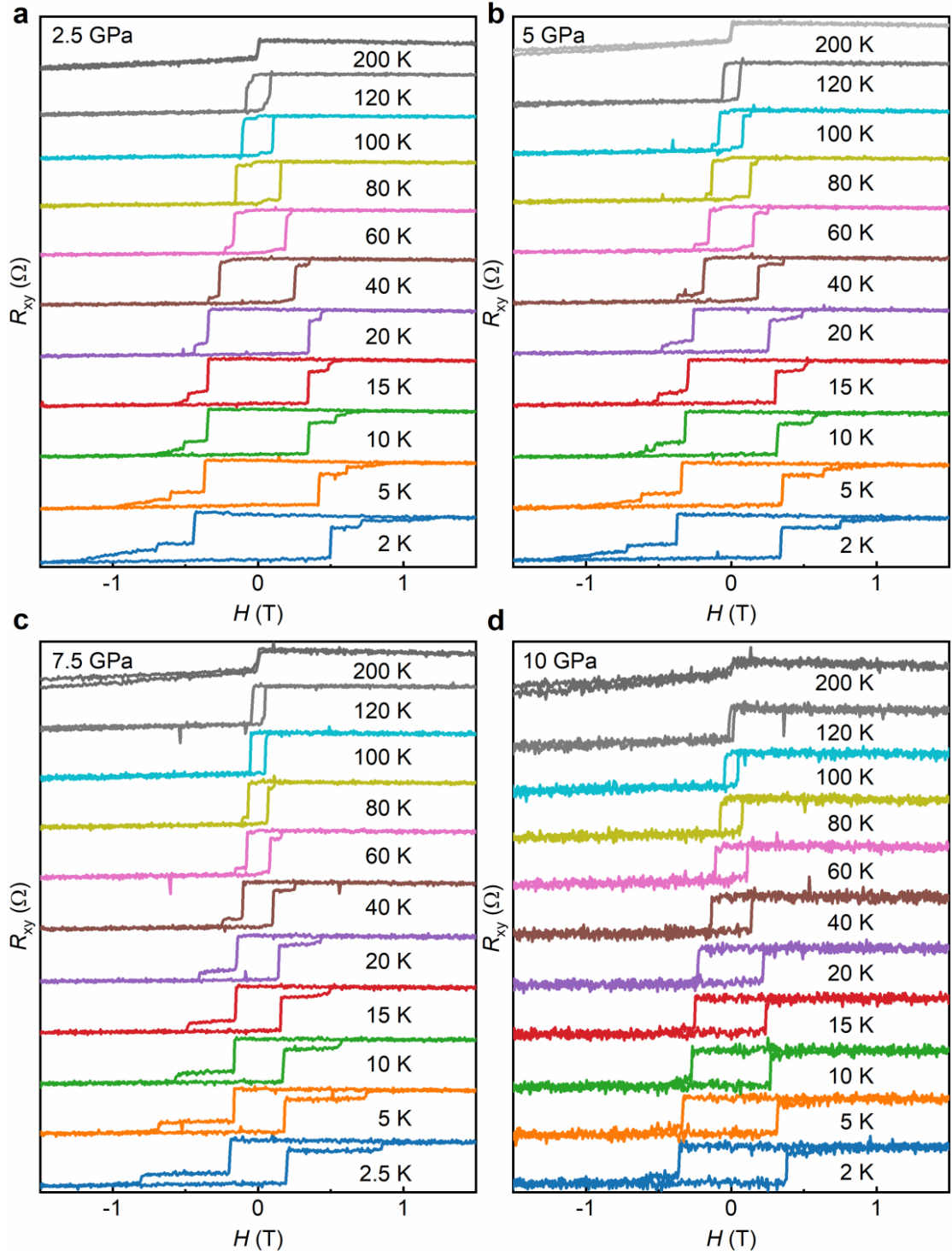

**Fig. S5.** $R_{xy}$-*H* **curves of the FGT/MPS heterostructure at different temperatures under various pressures.** $R_{xy}$ represents the transverse Hall resistance and *H* indicates the applied out-of-plane magnetic field. Each applied high pressure is 2.5 (a), 5 (b), 7.5 (c), and 10 (d) GPa, respectively.

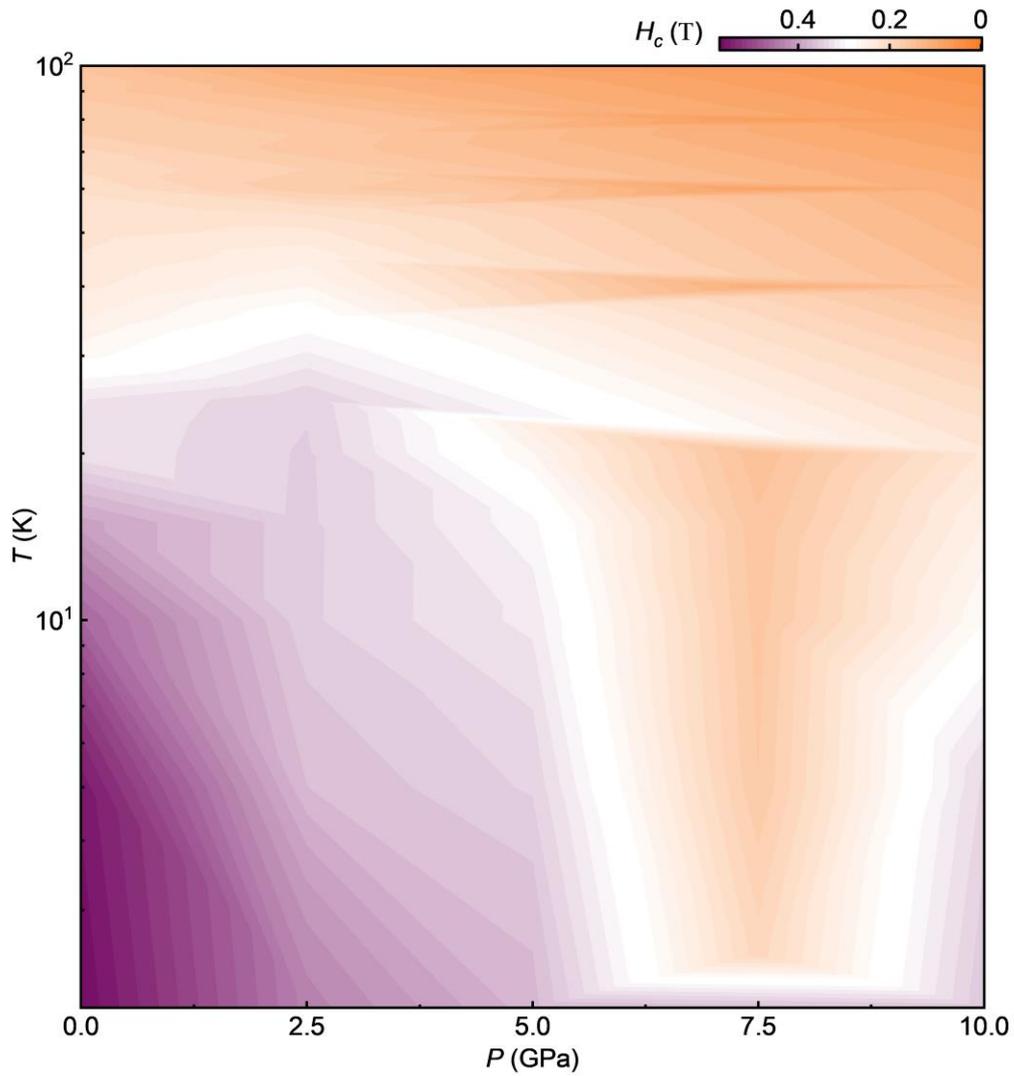

**Fig. S6. Temperature-pressure mapping of $H_c$ in FGT/MPS heterostructure.** $H_c$ increases as temperature reduces at a fixed pressure, normally indicating the ferromagnetic behavior of FGT. Meanwhile, at a fixed low temperature like 2 K, $H_c$ decreases then increases when the applied high pressure is increased to cross a critical pressure of ~7.5-10 GPa.

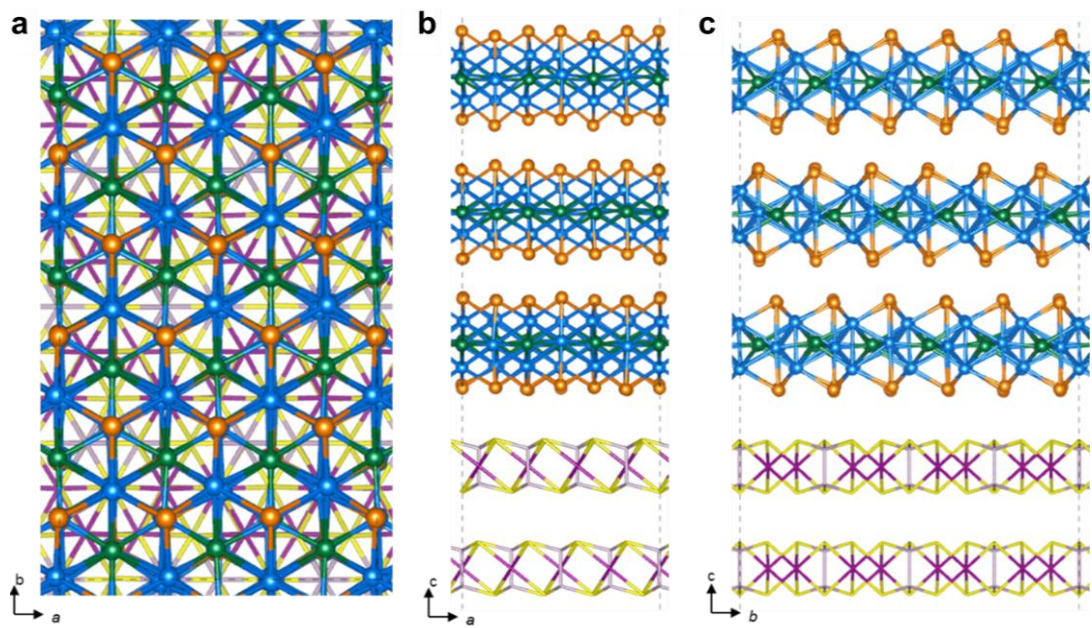

**Fig. S7. Top and side views of the 3×3 √3 3L-FGT/2×2 √3 BL-MPS heterostructure model.** Blue, green, brown, purple, pink and yellow balls represent the Fe atoms, Ge atoms, Te atoms, Mn atoms, P atoms and S atoms, respectively. The yellow and purple lines correspond to MPS substrate for clarity.

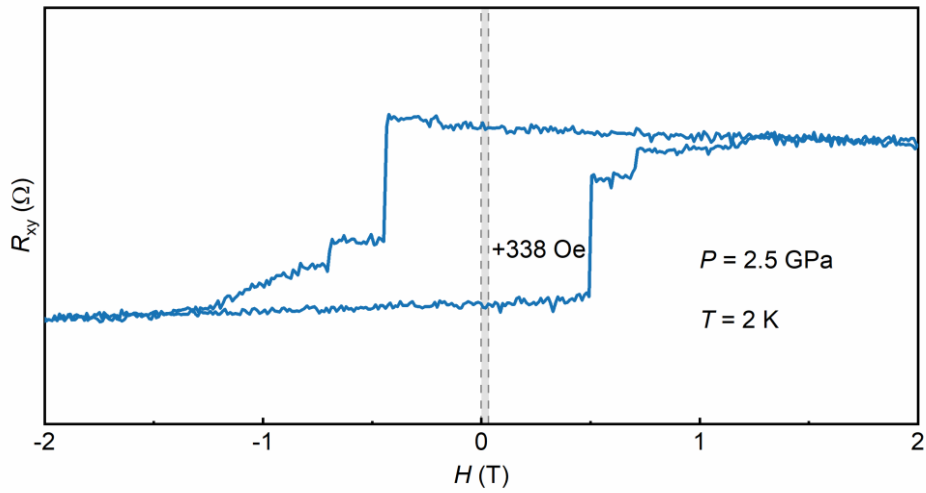

**Fig. S8. Certain exchange bias effect in the samples.** Similar exchange bias effects have been reported in many recent nice works. However, we should note that our central focus is the intermediate Hall plateau pinned by an emergent interfacial antiferromagnetism, instead of this commonly observed exchange bias effect.

**Supporting Tables**

| Stru. Config. | Mag. Config | | △E (meV/Fe) |
| --- | --- | --- | --- |
| | Intralyer | Interlayer | |
| Free standing 3L FGT | **FM** | **AAA** | **0.00** |
| | FM | ABA | 0.8 |
| | FM | ABB | 0.4 |
| | FM | AAB | 0.4 |
| | FiM1 | AAA | 103.4 |
| | AFM | AAA | 41.7 |
| 3L FGT under in-plane epitaxial strain | FM | AAA | 0.0 |
| | FM | ABA | 0.3 |
| | FM | AAB | 3.2 |
| 3L FGT on MPS | FM | AAA | 24.8 |
| | FM | ABA | 2.7 |
| | FM | ABB | 5.3 |
| | **FM** | **AAB** | **0.0** |
| | FiM1 | AAA | 152.4 |
| | FiM2 | AAA | 130.2 |
| 3L FGT on MPS under large pressure | FM | AAA | 0.0 |
| | FM | ABA | 10.7 |
| | FM | ABB | 6.0 |
| | FM | AAB | 5.84 |

**Table S1. Relative total energies of the free-standing and substrate-supported 3L FGT.** The charge transfer at the interface can modulate the inter-layer magnetic coupling in adjacent FGT layers, thereby stabilizing the FM-AAB magnetic structure as the ground state. To further investigate the impact of interfacial strain, we constructed a 3L FGT model under in-plane epitaxial strain to examine whether it would affect the magnetic ground states of the heterostructure. Our calculations reveal that while the substrate induced interfacial strain weakens the interlayer ferromagnetic coupling of FGT, it does not render the antiferromagnetic state as the ground state. This confirms that local charge transfer plays an important role in regulating the interlayer magnetic

ground state. Under applied pressure, the interlayer distance decreases, enhancing interfacial charge transfer and interfacial antiferromagnetic (AFM) interactions. Simultaneously, the intra-layer lattice constants and inter-layer distance variation induced by hydrostatic pressure also influence the magnetic ground state and magnetic anisotropy. Given the complexity of accurately characterizing and predicting the lattice structure changes and their correlation with external strain in heterojunction systems, theoretical predictions of the relationship between external pressure and magnetic transition remain challenging. We applied 4% compressive strain in the in-plane direction and 8% compressive strain in the interlayer direction to the FGT/MPS heterostructure model, with the objective of qualitatively evaluating the magnetic transition of the heterojunction under substantial strains. Interestingly, at a significantly large strain, theoretical calculations point to a strengthened interfacial FM interaction, which may account for the larger $H$c at 10 GPa for the FGT/MPS device. On the experimental point of view, the $H$c behavior above 10 GPa for FGT/MPS device is similar to that of pristine FGT nanoflake, making it difficult to testify the theory prediction at the large strain induced by a 10 GPa high pressure.